\documentclass[twocolumn,aps,prl,showpacs,superscriptaddress]{revtex4}
\usepackage{graphicx}
\newcommand{\GVB}{$G\left(V,B\right)$}
\begin{document}
\author{Yaroslav Tserkovnyak}
\author{Bertrand I. Halperin}
\affiliation{Lyman Laboratory of Physics, Harvard University,
Cambridge, Massachusetts 02138}
\author{Ophir M. Auslaender}
\author{Amir Yacoby}
\affiliation{Dept. of Condensed Matter Physics, Weizmann
Institute of Science, Rehovot 76100, Israel}

\title{Finite-size effects in tunneling between parallel quantum wires}

\begin{abstract}
We present theoretical calculations and experimental
measurements which reveal finite-size effects in the
tunneling between two parallel quantum wires, fabricated at
the cleaved edge of a GaAs/AlGaAs bilayer heterostructure.
Observed oscillations in the differential conductance, as a
function of bias voltage and applied magnetic field, provide
direct information on the shape of the confining potential.
Superimposed modulations indicate the existence of two distinct
excitation velocities, as expected from spin-charge separation.
\end{abstract}

\date{\today}


\maketitle

\begin{figure}[pb]
\includegraphics[width=3.0in,clip=]{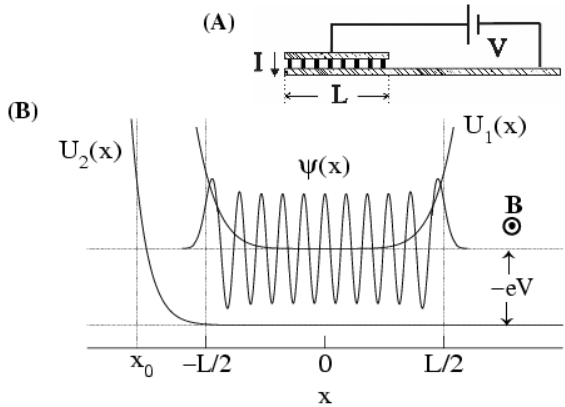}
\caption{\label{sc}Schematics of the circuit (A) and the
model (B). A wire of length $L$ runs parallel to a
semi-infinite wire. Boundaries of the wires are
formed by potentials $U_1(x)$ and $U_2(x)$.
Also drawn in (B) is a one-electron wave function $\psi(x)$
in the upper-wire confinement $U_1(x)$.
The energy and momentum of the tunneling electrons are governed
by voltage $V$ and magnetic field $B$.}
\end{figure}

One-dimensional (1D) electronic systems are a very fertile
ground for studying the physics of interacting many-body
systems. In one dimension, the elementary excitations are
collective spin and charge modes, the spectrum of which is
strongly influenced by the Coulomb interaction
\cite{Emery79}. An electron entering such a system must,
therefore, decompose into the corresponding eigenmodes,
resulting in a suppression of the tunneling density of
states. This suppression was detected in
a variety of experiments, such as tunneling from metal
contacts into carbon nanotubes \cite{Bockrath99} and resonant
tunneling in one dimension \cite{Ophir00}. A unique feature
of interacting electrons in one dimension, described by
Luttinger-liquid (LL) theory \cite{Emery79}, is the
decoupling of the spin and charge degrees of freedom, each of
which propagates with a different velocity determined by the
Coulomb interaction. To date, direct experimental
verification of this phenomenon is lacking. Moreover, issues
pertaining to the decoherence and relaxation of the
elementary excitations of the LL remain a challenge both
theoretically and experimentally.

Here we report a detailed experimental investigation and
theoretical explanation of a set of interference patterns in
the nonlinear tunneling conductance between two parallel
wires that were first reported in
Ref.~\cite{Auslaender:sc02}. A sketch of the tunneling
geometry in shown in Fig.~\ref{sc}(A). The interference appears
because the tunneling process is coherent to a very high
degree and is due to the finite length of the tunnel
junction. A wealth of information can be extracted from the
interference: The pattern itself encodes microscopic details
of the potentials in the wires, while the structure of its
envelope reflects the presence of two distinct excitation
velocities per electron mode in the data, as expected from
spin-charge separation. The decay of interference may also
yield information on decoherence processes of the elementary
excitations in 1D systems.

Fabrication of 1D quantum wires of exceptional quality has
been achieved by cleaved edge overgrowth in GaAs/AlGaAs
heterostructures \cite{Yacoby:prl96}, and much progress has
been made in experimental investigations of their transport
properties \cite{Picciotto:prl00,Auslaender:sc02}. Recently,
a method was developed to measure tunneling between two
parallel wires of this type \cite{Auslaender:sc02}: The wires
are formed at the cleaved edge of a wafer containing two
parallel quantum wells, only one of which is occupied by a
two-dimensional (2D) electron gas (2DEG) (cf.
Ref.~\cite{Auslaender:sc02} for details on the sample). A
voltage bias $V$ between the wires forces electrons to tunnel
through a narrow AlGaAs barrier separating them [see
Fig.~\ref{sc}(A)]. Measurements of the differential conductance
$G$ at $0.25$~K are made with standard lock-in techniques, as
a function of $V$ and $B$, a magnetic field applied
perpendicular to the plane containing the wires. This allows
to determine the complete dispersion relations of the
elementary excitations in the quantum wires
\cite{Auslaender:sc02}. Prominent features of the measured
\GVB\ can be understood in terms of a model which considers
two infinite parallel wires, and accounts for the
electron-electron interactions by means of LL theory
\cite{Emery79,Auslaender:sc02,Carpentier:cm01}. On the other
hand, an observed oscillation pattern results from the finite
length of the upper wire (UW). We show that key features of
the oscillations can be understood assuming that the ends of
the UW are defined by a soft confining potential, rather than
assuming sharp, square-well confinement. Most of our
discussion will employ a model of \emph{noninteracting}
electrons, which explains the most prominent features of the
interference patterns. Interaction effects will be discussed
at the end. Fig.~\ref{sc}(B) schematically shows the potentials
$U_1(x)$ and $U_2(x)$ felt by electrons in the wires.
Electrons in the UW are confined to a region of finite length
by top gates at both ends of the junction. Electrons in the
lower wire (LW) are reflected at the left end, but can pass
under the right-hand gate, rendering it semi-infinite. The
effective tunneling region is determined by the length of the
UW, which is approximately the region $|x| < L/2$ in
Fig.~\ref{sc}(B). The magnetic field gives a momentum boost
$\hbar Q=eBd$ to electrons tunneling from the UW to the LW,
$-e$ being the electron charge and $d$--the distance between
the centers of the wires; $V>0$ favors tunneling of electrons
from the LW to the UW.

\begin{figure}[pb]
\includegraphics[width=3.4in,clip=]{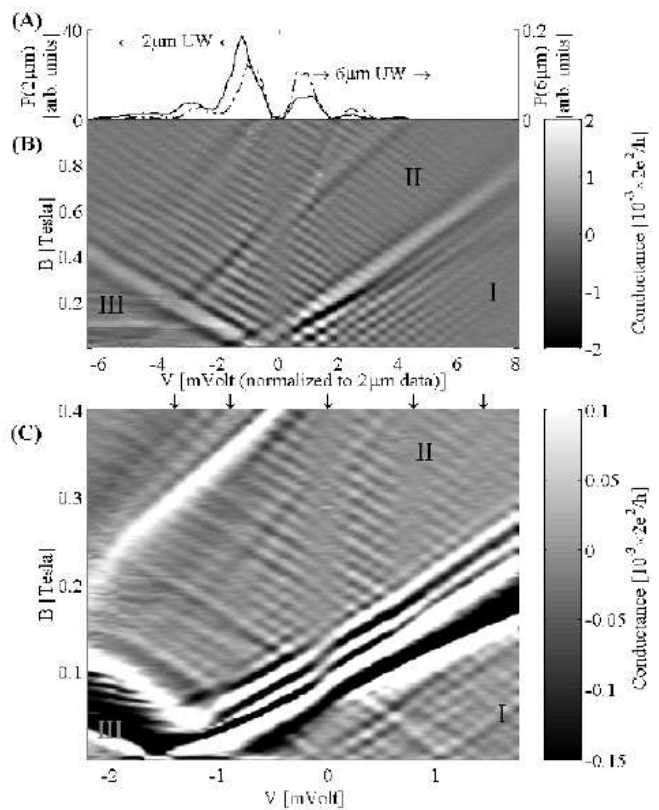}
\caption{\label{osc} Oscillations of \GVB\ at low field from
2~$\mu$m (B) and 6~$\mu$m (C) junctions. (A smoothed
background has been subtracted to emphasize the
oscillations.) The arrows in (C) mark the minima of the power $P$
of the oscillations in magnetic field as a function of the
voltage $V$. (A) shows $P$ for (B) (solid line) and (C)
(dashed), for which the abscissa is $2.3\times V$. The
brightest lines in (B) and (C), corresponding to tunneling between the
lowest modes, break the $V$-$B$ plain into regions I, II, and
III. Additional bright lines in II arise from other 1D
channels in the wires and are disregarded in our theoretical
analysis.}
\end{figure}

As observed in Ref.~\cite{Auslaender:sc02}, the differential
conductance has strong peaks along a set of curves in the
$V$-$B$ plane, where tunneling can occur with conservation of
energy and momentum between a Fermi point in one wire and an
electron mode in the other wire. Here we focus on the region
near the intersection of the dispersion peaks in \GVB\ at
zero magnetic field \cite{Auslaender:sc02} and voltage,
$V=(E_{\text{F}_2} - E_{\text{F}_1})/e$, necessary to
overcome the difference in Fermi energies $E_{\text{F}_1}$
and $ E_{\text{F}_2}$  of the UW and LW, respectively. (In
the experiments, several electron modes may be occupied in
the wires.  Here we consider only the mode in each wire with
the largest Fermi momentum along the wire.) The most
spectacular manifestation of the breaking of translational
invariance is the appearance of a regular pattern of
oscillations away from the dispersion curves. Fig.~\ref{osc}
shows typical examples of the patterns measured at low
magnetic field. The lines that correspond to the dispersions
appear as pronounced peaks that extend diagonally across the
figure. Additionally, we observe numerous secondary peaks
running parallel to the dispersions. These side lobes are
asymmetric: They always appear to the right of the principal
dispersion peaks. The result is a checkerboard of
oscillations in region I, stripes in region II, and no
regular pattern in region III (see Fig.~\ref{osc}).
When the lithographic length $L$ is increased from 2~$\mu$m [cf.
Fig.~\ref{osc}(B)] to 6~$\mu$m [cf. Fig.~\ref{osc}(C)], the
frequency in $V$ and $B$ increases by a factor of $\approx3$.
The period is approximately related to $L$, $d$, and the
Fermi velocity $v_{\text{F}}$ by $|\Delta
V|L/v_{\text{F}}=|\Delta B|Ld=2\pi\hbar/e$.
Upon close examination of Figs.~\ref{osc}(B) and \ref{osc}(C),
one can discern a modulation of the
interference that causes a series of faint streaks parallel to the $B$-axis,
where the oscillations are suppressed. The strength of this modulation is
shown in Fig.~\ref{osc}(A). As explained below, this is
a moire pattern created by two superimposed sets of
interference, each resulting from a distinct velocity that is
present in the data.

We base our theoretical analysis on a phenomenological
tunneling Hamiltonian
\begin{equation}
H_{\text{tun}}=T\sum_s\int
dx\Psi^\dagger_{s1}(x)\Psi_{s2}(x)e^{-iQx}+\text{H.c.}\,,
\end{equation}
where $\Psi_{si}(x)$ is the spin-$s$ electron field operator
for the $i$th wire ($i=1$ denotes the UW and $i=2$ the LW).
Since the Zeeman energy in GaAs is small, we ignore the spin
degrees of freedom, and characterize the electrons in the UW
by a discrete set of energy eigenstates $\psi_n (x)$. The
eigenstates in the lower wire form a continuum, which we
write as $\varphi_{k_2} (x)$, indexed by wave vector $k_2$.
Treating tunneling to the lowest nonvanishing order in
perturbation, we find for the current
\begin{equation}
I\propto\text{sgn}(V)|T|^2\sum_m |M(n,Q,V)|^2 \,, \label{I1b}
\end{equation}
where $n = n_{\text{F}} +\text{sgn}(V)m$,  with
$n_{\text{F}}$ being the index of the state $\psi_n$ just
below $E_{\text{F}_{1}}$; $\sum_m$ is a sum over integers $m$
with $[\text{sgn}(V)-1]/2<m<e|V|L/\pi\hbar v_{\text{F}}$; and $M$
is the tunneling matrix element between state $\psi_n$ and
state $\varphi_{k_2}$, the energy of which is lower by $eV$.
Specifically, $M$ is given by
\begin{equation}\
M= \int dx \psi_n^* (x) e^{-iQx} \varphi_{k_2} (x) \,.
\label{Mb}
\end{equation}
If $e|V|$ is not too large, we can linearize about the Fermi
wave vectors $k_{\text{F}_{i}}$, and $k_2$ is then given by
$(k_2-k_{\text{F}_2})v_{\text{F}_2}=\text{sgn}(V)v_{\text{F}_1}m\pi/L-
eV/\hbar$.

As a starting point, we consider infinite square-well
confinement in the region $|x|<L/2$ of the UW, so that
$\psi_n \propto \sin n\pi(x/L-1/2)$. We also assume that the
potential in the LW is infinite for $x < x_0$, so that the
states in the LW have the form $\varphi_{k_2}\propto\sin
k_2(x-x_0)$. In line with the experiments, we assume that
$k_{\text{F}_{1,2}}$ differ slightly by $\Delta
k_{\text{F}}=k_{\text{F}_{1}}-k_{\text{F}_{2}}$, but neglect
the difference in the $v_{\text{F}}$'s of the two wires,
which is only a few percent \cite{Auslaender:sc02}. As a
result, $|M|^2$ is independent of the index $n$ and is given
by $|M|^2=|M^{(+)}|^2+|M^{(-)}|^2$, where
$M^{(\pm)}=\sin[\kappa^{(\pm)}L/2]/[\kappa^{(\pm)}L/2]$ and
$\kappa^{(\pm)}=\Delta k_{\text{F}}+eV/\hbar v_{\text{F}}\pm
Q$. (In the limit $L\to\infty$, the two terms $|M^{(\pm)}|^2$
become $\delta$-functions.) Similar results were obtained by
\citet{Boese:prb01} who considered tunneling between two
infinite noninteracting wires through a window of finite
length $L$ \cite{Governale:prb00}.

Differentiating Eq.~(\ref{I1b}) to obtain \GVB, we find that
whenever the applied voltage matches a discrete energy level
of the UW with the Fermi level of the LW, the sum in
Eq.~(\ref{I1b}) exhibits a step, yielding a series of
$\delta$-function peaks in the conductance. These peaks are
not seen in the experiments, possibly because of smearing due
to finite temperature and/or 1D-2D scattering in the UW. We
may therefore disregard the discreteness of the sum and write
$I \propto V \,|M(Q,V)|^2$, which gives a pronounced
oscillatory contribution $\propto V\partial |M|^2 /\partial
V$ to $G$. (When $L\to\infty$, this oscillation pattern
disappears and the conductance becomes $G\propto|M(Q,V)|^2$
resulting in bright peaks along the dispersion curves.) There
are several features in the data, most notably the
interference side-lobe asymmetry, not captured by this
idealized model. The asymmetry of side lobes can be well
understood if, instead of assuming square-well confinement,
we consider a smooth potential well $U_1(x)$, giving rise to
a WKB wave function form for the electronic states in the UW
[see Fig.~\ref{sc}(B)].

\begin{figure}[pt]
\includegraphics[width=3.2in,angle=0,clip=]{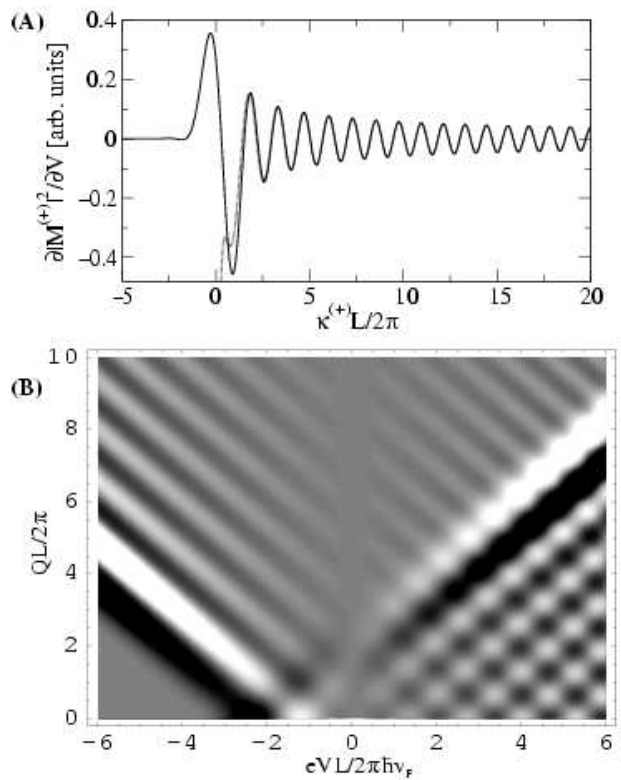}
\caption{\label{as} (A) $\partial|M^{(+)}|^2/\partial V$,
obtained numerically using the WKB wave function $\psi_{100}(x)$
and $\alpha=6$ in Eq.~(\ref{U}). The
dotted curve shows an approximation [Eq.~(\ref{Mcos})].
(B) Calculated oscillation pattern
$G\propto V\partial|M|^2/\partial V$ using the numerical result from
Fig.~\ref{as}(A).
$v_{\text{F}_1}=v_{\text{F}_2}=v_{\text{F}}$ and $\Delta
k_{\text{F}}=4\pi/L$.}
\end{figure}

As an example, we model the UW by a symmetric potential of
the form
\begin{equation}
U_1(x)=E_{\text{F}_1}|2x/L|^\alpha\,. \label{U}
\end{equation}
The limit $\alpha\to\infty$ recovers the case of a square
quantum well while a finite value of $\alpha\ge2$ defines
smooth walls. The appropriate choice of $\alpha$ should
increase with $L$. As a specific example, we consider the
case $\alpha=6$. We assume that the boundaries are soft
enough so that one can use a WKB approximation for wave
function $\psi_n$ in Eq.~(\ref{Mb}). We set
$n_{\text{F}}=100$ for the state at the Fermi energy in the
UW, which corresponds to electron density of
$\approx100~\mu$m$^{-1}$ \cite{Auslaender:sc02} in the most
occupied 1D channel, with $L=2$ $\mu$m. The theory again
gives two separate contributions $|M^{(\pm)}|^2$ to $|M|^2$,
which are functions of the variables $\kappa^{(\pm)}$, as
before. In Fig.~\ref{as}(A), we plot the results of a numerical
calculation of $\partial |M^{(+)}|^2/
\partial V$ as a function of $\kappa^{(+)} $. Unlike the
square-well case, the smooth boundaries lead to a very
asymmetric oscillation pattern for positive and negative
$\kappa^{(+)}$. Furthermore, the amplitude of the
oscillations in Fig.~\ref{as}(A) drops much slower than
in the case of infinitely
steep walls. Results of a numerical calculation of \GVB\ for
the soft-confinement model, obtained from Eq.~(\ref{Mb}), are
displayed in Fig.~\ref{as}(B). They are very similar, at least
qualitatively, to Fig.~\ref{osc}(B).

The numerical results can be understood analytically as
follows. To find $M^{(+)}$, we substitute $e^{ik_2 x}$ for
$\varphi_{k_2}$ in Eq.~(\ref{Mb}), and replace $\psi_n$ by
the right-moving WKB component $k(x)^{-1/2}e^{i
k_{\text{F}_1}x}e^{-is(x)}$, where
$k(x)=k_{\text{F}_1}[1-U_1(x)/E_{\text{F}_1}]^{1/2}$ and
$s(x)=\int_0^xdx^\prime [k_{\text{F}_1} -  k(x^\prime)]$.
Using $k_2-Q=k_{\text{F}_1}-\kappa^{(+)}$, we see that if $0
<\kappa^{(+)} <k_{\text{F}_1}$, there are two points, $x^+>0$
and $x^-<0$, where $k(x)=k_2-Q$ and the integrand in
Eq.~(\ref{Mb}) has a stationary phase.  We evaluate the
integral near these two points using the method of steepest
descents, and add the results to obtain an estimate of
$M^{(+)}$. In the case of a symmetric potential,
$U_1(x)=U_1(-x)$, one finds
\begin{equation}
M^{(+)}\approx\sqrt{\frac{16\pi E_{\text{F}}k^{-2}_{\text{F}}}
{U_1^\prime(x^+)}}
\cos\left[\kappa^{(+)}  x^+ -s(x^+)-\frac{\pi}{4}\right]\,.
\label{Mcos}
\end{equation}
In Fig.~\ref{as}(A), we plot $\partial|M^{(+)}|^2/\partial V$,
obtained using Eq.~(\ref{Mcos}) as a dotted curve: it is in
an excellent agreement with the full numerical calculation
(solid line) for large positive $\kappa^{(+)}$. The
stationary phase approximation (SPA) is bad for small values
of $\kappa^{(+)}$, where the conditions for its validity are
violated. For the potential of Eq.~(\ref{U}), $2x^+/L \approx
[2 \kappa^{(+)}/k_{\text{F}}]^{1/\alpha}$ for
$\kappa^{(+)}\ll k_{\text{F}}$, and the phase $[\kappa^{(+)}
x^+ - s(x^+)]$ in Eq.~(\ref{Mcos}) is equal to $\kappa^{(+)
}x^+ \alpha / (\alpha + 1)$. For negative $\kappa^{(+)}$, the
equation $k(x)=k_2 -Q$ does not possess a real-valued
solution, but after deforming the contour of integration
[Eq.~(\ref{Mb})] into the complex plane, we obtain a
complex-valued solution
$2x/L\approx(2|\kappa^{(+)}|/k_{\text{F}})^{1/\alpha}e^{-i\pi/\alpha}$.
The argument of the cosine in Eq.~(\ref{Mcos}) then has an
imaginary part, which causes the magnitude of $|M^{(+)}|^2$
to rapidly fall by a factor of $\approx e^{-2\pi^2/\alpha}$
over each period of oscillation. The approximation
[Eq.~(\ref{Mcos})] can be used to predict both the period of
the conductance oscillations and their amplitude. The period
of the oscillations in the case of smooth confinement is
given by $\Delta\kappa^{(+)}\approx2\pi/(x^+-x^-)$, rather
than $\Delta \kappa^{(+)}=2\pi/L$ for the square-well
potential. In particular, this means that at low
$\kappa^{(+)}$ the period can be significantly larger than
the value $2\pi/L$, which it approaches for large
$\kappa^{(+)}$. The prefactor $1/U_1^\prime(x^+)$ in $|M|^2$
falls off less rapidly with $\kappa^{(+)}$ in the case of a
smooth potential than the $1/[\kappa^{(+)}]^2$ dependence for
the square well. For Eq.~(\ref{U}), we find
$1/U_1^\prime(x^+)\propto[\kappa^{(+)}]^{1/\alpha-1}$, which
goes only as $1/\kappa^{(+)}$ for large $\alpha$ or
$1/\sqrt{\kappa^{(+)}}$ for $\alpha=2$. (Because of
the WKB approximation, we cannot recover the square-well
result by setting $\alpha\rightarrow\infty$ in our
expressions.) A larger period for small $ \kappa^{(+)}  $ and
relatively slow fall-off of the amplitude with increasing $
\kappa^{(+)} $ are both qualitatively consistent with the
experimental results.

An analysis using the SPA and WKB approximations can also be
applied to an \emph{interacting} electron system in a pair of
wires with soft confinement. The conductance is determined by
Green's function
$G_{12}=\langle[\Psi^\dagger_{s2}\Psi_{s1}(x,t),
\Psi^\dagger_{s^\prime1}\Psi_{s^\prime2}(x^\prime,0)]\rangle$.
The right- and left-mover contributions to $G_{12}$ can be
approximated by the respective contributions $C_{R,L}^\infty$
for an infinite wire, given by LL theory
\cite{Carpentier:cm01}, multiplied by phase factors of the
form $e^{\pm i [s(x) - s(x^\prime)]}$.  For a pair of
coupled, nearly identical wires, the leading singularities in
$C_{R,L}^\infty$ occur at velocities $v_c$ and $v_s$,
corresponding to charge and spin excitations that have
\emph{opposite sign in the two wires}, as the symmetric modes
are not excited in the tunneling process.

We find that LL theory preserves the key qualitative features
in Fig.~\ref{as}(B), such as the asymmetry of the interference
pattern, but brings about additional features resulting from
the presence of two distinct velocities. In particular, the
main dispersion peaks split into two lines with slopes
defined by $v_c$ and $v_s$. As a consequence, the overlap of
the interference side lobes formed along the dispersion
slopes creates a moire pattern resulting in a periodic
modulation of the conductance oscillations along the voltage
axis. The distance between the corresponding stripes of suppressed
conductance running parallel to the field axis is $\Delta
V_{\text{slow}} \approx 2\pi\hbar v_c v_s/ eL(v_c-v_s)$,
which is larger than the period of oscillations, $\Delta
V_{\text{fast}} \approx 4\pi\hbar v_c v_s/eL(v_c+v_s)$. Such
a beating phenomenon in the interference is seen in
Fig.~\ref{osc} and may thus be a direct consequence of
spin-charge separation in one dimension.
For the 2~$\mu$m wire, from the data in Figs.~\ref{osc}(A),(B),
we estimate $\Delta V_{\text{slow}}=1.8\pm0.4$~mV and $\Delta
V_{\text{fast}}=0.7\pm0.1$~mV, and then find $v_s/
v_c=0.67\pm0.07$. This result agrees with estimates made
previously in a different regime \cite{Auslaender:sc02}. A
detailed comparison between theory and experiment can thus be
used to study electron-electron interactions as well as the
shape of the confining potentials along the quantum wires.
Another effect predicted by LL theory, and present in the
data, is a decrease in the value of $G$ at low bias, i.e.,
zero-bias anomaly. In the limit $L\to\infty$, the
interference side lobes disappear, and what remains for the
interacting wires are the split dispersion peaks, given by
the LL theory of Ref.~\cite{Carpentier:cm01}. A more thorough
discussion of LL theory for the double-wire system will be
given elsewhere \cite{Tserkovnyak:prep}.

We have enjoyed illuminating discussions with B. Kramer, Y.
Oreg, M. Sassetti, and A. Stern. This work was supported in
part by NSF Grant DMR 99-81283, the Schlumberger Foundation,
the US-Israel BSF,
% and by a research grant from
the Fusfeld Research Fund,
%OMA is supported by a grant from
and the Israeli Ministry of Science.

\end{document}